\documentclass[11pt,a4paper,aps,amsmath,amssymb,superscriptaddress,reprint,secnumarabic, graphics,floatfix,nofootinbib,
tightenlines,nobibnotes,aps,prd]{revtex4-1}
\usepackage[paperwidth=210mm,paperheight=297mm,centering,hmargin=2cm,vmargin=2.5cm]{geometry}
\usepackage{color}
\usepackage[table]{xcolor}
\usepackage{tabu}
\usepackage{array}
\newcommand{\rank}[4]{{#1}^{#2}{}_{{#3} {#4}}}
\newcommand{\ranks}[5]{{#1}^{#2}{}_{{#3}{#4}{#5}}}
\newcommand{\apvec}[2]{\underset{\fontsize{3}{2}\selectfont {#1}}{#2}{}}

\begin{document}
     \title{A Single Equation Of Gravity And Electromagnetism On Parallelizable Manifold Using Dolan-McCrea Variational Method}
     \thanks{Also christened: Absolute Parallelism}
     \author{Christian Nwachioma}
     \email{chris.nwachioma@gmail.com}
     \author{Farida Tahir}    
     \email{farida\_tahir@comsats.edu.pk}
     \affiliation{COMSATS Institute of Information Technology, Islamabad}
     %\author{Shi-Hai Dong}
     %\email{dongsh2@yahoo.com}
     \affiliation{National Mathematical Center, Abuja}
     
     \begin{abstract}
         The crucial but undocumented Dolan-McCrea variational method is richly applied. Using the said method, we analytically derived a field equation comprising entirely of geometric structures and we investigate how effectively it describes gravitational and electromagnetic phenomena. The procedure we adopted involved constructing a scalar invariant as was the case for Einstein's General Relativity (GR) except that the scalar of parameterized Absolute Parallelism geometey consists of the Ricci scalar plus an additional term, which is essentially the contortion. %Based on earlier discourse, a good starting point for this section would be the connections of the PAP-space.
    \end{abstract}       
    \maketitle
    
    It is known that parallelizable or absolute parallelism (AP-)manifold has defined on it, multiple connections with the basic ones being the canonical connection, the dual of the canonical connection, the symmetric connection and also the Levi-Civita connection\cite{Wanas2002,Nwachioma2017}. It has been shown that the curvatures with respect to any linear connection on the parallelizable manifold can be written as products of torsion\cite{YOUSSEF2007}; this implies that vanishing of the torsion results in identical vanishing of the various curvatures on the manifold making the manifold seems flat. However, there's a way to walk around this issue as suggested in \cite{Wanas2000}. The problem has been remedied by adopting the parameterized AP (PAP-)space or by using the W-tensor\cite{Wanas2016}. The W-tensor has been used to formulate a unified field equation of gravity and electromagnetism\cite{WanasOsmanKholy2015,Wanas2015}. Reinventing the wheel is not what we are set to achieve here, instead, we are presenting a rich mathematical insight by revitalizing the technically moribund Dolan-McCrea variational principle and applying it toward arriving at a purely geometric unified field equation of gravity and electromagnetism. To do so, we use a PAP connection, namely the dual of the connection stated below\cite{Wanas2002}.
    %By default, the symmetric connection of the PAP space has zero torsion and thus not a good candidate for unification of a symmetric and %a skew-symmetric theories. The curvature is however, non-vanishing.
    %The parameterized connections have the advantage over the conventional (AP-space) canonical connection in that it has simultaneous %non-vanishing curvature and torsion. This connection is denoted as follows\cite{Wanas2002}.
    \begin{equation}
    \label{eqn6-1::1}
    \nabla^\alpha{}_{\mu\nu}=\{^\alpha_{\mu\nu}\}+q\gamma^\alpha{}_{\mu\nu}
    \end{equation} 
    where `q' is a scalar parameter; $\{^\alpha_{\mu\nu}\}$ is the Levi-Civita connnection, and $\gamma^\alpha{}_{\mu\nu}$ is the spacetime contortion. The daul of Eq.\eqref{eqn6-1::1} is given by:%Recall that by default, the canonical connection of the conventional AP-space has vanishing curvature but non-vanishing torsion. %The parameterized canonical connection \eqref{eqn6-1::1} will reduce to the Riemannian connection whenever q=0: $\nabla^\alpha{}_{\mu\nu}=\{^\alpha_{\mu\nu}\}$ and will degenerate to the convention AP-space canonical connection whenever q=1: $\nabla^\alpha{}_{\mu\nu}=\Gamma^\alpha{}_{\mu\nu}$. This means that the situation where q=1 will present a space with zero curvature (but nonzero torsion). We can make a restriction that q never equals unity and go ahead to formulate a `unified' field theory using the parameterized canonical connection.But why make an undue and unhealthy restriction on q when we have a structure that can allow us to be carefree. So, in this attempted formulation of a unified field theory, we would like to clinch to the PAP structure called the parameterized dual connection given by \eqref{eqn6-1::2}.
    \begin{equation}
    \label{eqn6-1::2}
    \tilde{\nabla}^\alpha{}_{\mu\nu}=\nabla^\alpha{}_{\nu\mu}.
    \end{equation}
    Eq.\eqref{eqn6-1::2} has the advantage that even though the parameter $q=1$, we do not get simultaneous vanishing of the curvature and torsion tensors. And setting $q=0$ switches off electromagnetic phenomena and reduces the system to the Riemannian situation.
    \\[0.2cm]
    \section{Curvature With Respect To The Dual PAP Connection}
    Let \textbf{D} be the curvature tensor corresponding to the parameterized dual connection of the PAP-space. In consonance with the Riemann curvature tensor, \textbf{D} can be constructed as follows:
    \begin{equation}
    \begin{split}
    D^\alpha{}_{\mu\nu\sigma}=\tilde{\nabla}^\alpha{}_{\mu\sigma,\nu}-\tilde{\nabla}^\alpha{}_{\mu\nu,\sigma}+\tilde{\nabla}^\epsilon{}_{\mu\sigma}\tilde{\nabla}^\alpha{}_{\epsilon\nu}-\tilde{\nabla}^\epsilon{}_{\mu\nu}\tilde{\nabla}^\alpha{}_{\epsilon\sigma}\label{eqn6-1::3a},
    \end{split}
    \end{equation}
    where \textit{comma} denotes ordinary differentiation with respect to the indicated spacetime axes and \textit{semicolon} shall denote covariant differentiation with respect to the Levi-Civita connection. Putting Eq.\eqref{eqn6-1::2} into Eq.\eqref{eqn6-1::3a}, we have:
    \begin{widetext}
    \begin{equation}
    \begin{split}
       \ranks{D}{\alpha}{\mu}{\nu}{\sigma}=\{^\alpha_{\sigma\mu}\}_{,\nu}-\{^\alpha_{\nu\mu}\}_{,\sigma}+\{^\epsilon_{\sigma\mu}\}\{^\alpha_{\nu\epsilon}\}-\{^\epsilon_{\nu\mu}\}\{^\alpha_{\sigma\epsilon}\}+q\big[\ranks{\gamma}{\alpha}{\sigma}{\mu}{,\nu}-\ranks{\gamma}{\alpha}{\nu}{\mu}{,\sigma}\\+\rank{\gamma}{\epsilon}{\sigma}{\mu}\{^\alpha_{\epsilon\nu}\}-\rank{\gamma}{\alpha}{\sigma}{\epsilon}\{^\epsilon_{\mu\nu}\}-\rank{\gamma}{\alpha}{\epsilon}{\mu}\{^\epsilon_{\sigma\nu}\}+\rank{\gamma}{\alpha}{\epsilon}{\mu}\{^\epsilon_{\sigma\nu}\}-\rank{\gamma}{\epsilon}{\nu}{\mu}\{^\alpha_{\epsilon\sigma}\}\\+\rank{\gamma}{\alpha}{\nu}{\epsilon}\{^\epsilon_{\mu\sigma}\}+q(\rank{\gamma}{\epsilon}{\sigma}{\mu}\rank{\gamma}{\alpha}{\nu}{\epsilon}-\rank{\gamma}{\epsilon}{\nu}{\mu}\rank{\gamma}{\alpha}{\sigma}{\epsilon})\big]\\=\ranks{R}{\alpha}{\mu}{\nu}{\sigma}+q\big[ \ranks{\gamma}{\alpha}{\sigma}{\mu}{;\nu}-\ranks{\gamma}{\alpha}{\nu}{\mu}{;\sigma}+q(\rank{\gamma}{\epsilon}{\sigma}{\mu}\rank{\gamma}{\alpha}{\nu}{\epsilon}-\rank{\gamma}{\epsilon}{\nu}{\mu}\rank{\gamma}{\alpha}{\sigma}{\epsilon})\big]
    \label{eqn6-1::3b}.	
    \end{split}
    \end{equation}
    \end{widetext}
    %\subsection{Contraction of $\ranks{D}{\alpha}{\mu}{\nu}{\sigma}$}
    To obtain the Ricci tensor analogue, we set $\sigma=\alpha$ while noting that the contortion, $\mathbf{\gamma}$ is skew-symmetric with respect to the first pair of indices and symmetric with respect to the last pair of indices and the Riemannian curvature tensor is skew-symmetric with respect to both the first pair and second pair of indices. With this in mind, we put away terms with vanishing results in the resultant $(0,2)$ tensor field $D^\alpha{}_{\mu\nu\alpha}$.  And the result of this contraction is:
    \begin{widetext}
    \begin{eqnarray}
    \begin{split}    
        D_{\mu\sigma}&=\ranks{D}{\alpha}{\mu}{\nu}{\alpha}\\&=\ranks{R}{\alpha}{\mu}{\nu}{\alpha}+q\big[ \ranks{\gamma}{\alpha}{\alpha}{\mu}{;\nu}-\ranks{\gamma}{\alpha}{\nu}{\mu}{;\alpha}+q(\rank{\gamma}{\epsilon}{\alpha}{\mu}\rank{\gamma}{\alpha}{\nu}{\epsilon}-\rank{\gamma}{\epsilon}{\nu}{\mu}\rank{\gamma}{\alpha}{\alpha}{\epsilon})\big]\\&=R_{\mu\nu}+q(-\ranks{\gamma}{\alpha}{\nu\mu}{;\alpha}+q\rank{\gamma}{\epsilon}{\alpha}{\mu}\rank{\gamma}{\alpha}{\nu}{\epsilon})
    \label{eqn6-1::3c}.
    \end{split}	    
    \end{eqnarray}
    \end{widetext}
    %%%%%%%%%%%%%%%%%  SECOND FROMATTING STARTS  %%%%%%%%%%%%%%%%%%%%
    %\subsection{Scalar curvature of the PAP space}
    Next, we sum over the diagonal entries of Eq.\eqref{eqn6-1::3c}. Let $D$ be the result of the sum and it's called the scalar curvature presented as:
    \begin{equation}
    \begin{split}
    D&=g^{\mu\nu}D_{\mu\nu}\\&=g^{\mu\nu}\big[R_{\mu\nu}+q(-\ranks{\gamma}{\alpha}{\nu\mu}{;\alpha}+q\rank{\gamma}{\epsilon}{\alpha}{\mu}\rank{\gamma}{\alpha}{\nu}{\epsilon})\big]\\&=R-qC^\alpha{}_{;\alpha}+q^2\gamma^\epsilon{}_{\alpha\mu}\gamma^{\alpha\mu}{}_\epsilon\label{eqn6-1::4a}.			
    \end{split}
    \end{equation}
    
    \section{Field Equations}
    Along the path of stationary action, the term involving covariant differentiation of the basic form $c^\alpha$ will not contribute to the variation; so we may write Eq.\eqref{eqn6-1::4a} as: 
    \begin{equation}
    \begin{split}
    D&=R+q^2\gamma^\epsilon{}_{\alpha\mu}\gamma^{\alpha\mu}{}_\epsilon\label{eqn6-1::5a},
    \end{split}			
    \end{equation}
    where R represents the Ricci scalar and $Q=1/2q^2\Lambda^{\epsilon\alpha\mu}\gamma_{\alpha\mu\epsilon}$. Physically, the scalar curvature, $D$ can represent a Lagrangian density having a negative weight; so, we have to multiply it by a scalar capacity $\chi$ with weight of equal magnitude but of opposite sign. Then the absolute invariant Lagrangian, $L$ is defined as: 
    \begin{equation}
    \begin{split}
    L&:=\chi D\\&=\chi (R+Q)\label{eqn6-1::5b}; \qquad \chi=det(\apvec{i}{\chi}_\rho).
    \end{split}
    \end{equation}
    The Lagrangian density is in general a function of the parallelization vector, therefore Variation of $L$ with respect to the parallelization vector is studied upto second derivative. But since the second derivative contributes not to the variational principle, we need not evaluate terms involving second derivative of $\apvec{i}{\chi}$.    
    %%%Variation of L with respect to the parallelization vector $\apvec{i}{\chi}_\rho$ is given in Eq.\eqref{eqn6-1::5c} up to term %%involving %%second derivative of the parallelization vector $\apvec{i}{\chi}$ because the Lagrangian density is in general a function %%%of the parallelization vector, its first and its second derivatives. But since the second derivative contributes not to the %%variational principle, we need not evaluate terms involving 
    
    For Eq.\eqref{eqn6-1::5b}, the Euler-Lagrange equation is given by:
    \begin{equation}
    \begin{split}
    %0&=\frac{\delta D}{\delta\apvec{i}{\chi_\rho}}\apvec{i}{\chi}_\nu\\
    %&=\frac{1}{\chi}\Bigg[ \frac{\partial L}{\partial \apvec{i}{\chi_\rho}}-\Big(\frac{\partial L}{\partial \apvec{i}{\chi}_{\rho,\eta}}\Big)_{,\eta}+\Big(\frac{\partial L}{\partial\apvec{i}{\chi}_{\rho,\eta\tau} }\Big)_{,\tau\eta}   \Bigg]\apvec{i}{\chi}_\nu\\	
    0&=\frac{1}{\chi}\Bigg[ \frac{\partial R}{\partial \apvec{i}{\chi_\rho}}-\Big(\frac{\partial R}{\partial \apvec{i}{\chi}_{\rho,\eta}}\Big)_{,\eta}+\Big(\frac{\partial R}{\partial\apvec{i}{\chi}_{\rho,\eta\tau} }\Big)_{,\tau\eta}   \Bigg]\apvec{i}{\chi}_\nu\\
    &+\frac{1}{\chi}\Bigg[ \frac{\partial Q}{\partial \apvec{i}{\chi_\rho}}-\Big(\frac{\partial Q}{\partial \apvec{i}{\chi}_{\rho,\eta}}\Big)_{,\eta}+\Big(\frac{\partial Q}{\partial\apvec{i}{\chi}_{\rho,\eta\tau} }\Big)_{,\tau\eta}\Bigg]\apvec{i}{\chi}_\nu\label{eqn6-1::5c}. 
    \end{split}
    \end{equation}

    It is well known the first part of Eq.\eqref{eqn6-1::5c} or the part involving the Ricci scalar will give the Einstein tensor\cite{Harko2011}. Presently, we must concentrate on the second part.
    
    \begin{equation}
    \begin{split}
    \frac{\partial Q}{\partial\apvec{i}{\chi}_\rho}&=\frac{\partial}{\partial\apvec{i}{\chi}_\rho}(1/2q^2\chi\Lambda^{\epsilon\alpha\mu}\gamma_{\alpha\mu\epsilon})\\&=\frac{1}{2}q^2\Big(\frac{\partial \chi}{\partial\apvec{i}{\chi}_\rho}\Lambda^{\epsilon\alpha\mu}\gamma_{\alpha\mu\epsilon}+\chi\frac{\partial\Lambda^{\epsilon\alpha\mu}}{\partial\apvec{i}{\chi}_\rho}\gamma_{\alpha\mu\epsilon}\\&\qquad+\chi\Lambda^{\epsilon\alpha\mu}\frac{\partial\gamma_{\alpha\mu\epsilon}}{\partial\apvec{i}{\chi}_\rho}\Big)\label{eqn6-1::5d},
    \end{split}
    \end{equation}
    %%%%%%%%%%%%%%%%%%%%%%%%%%%%%%%% THIRD FORMATTING STARTS %%%%%%%%%%%%%%%%%%%%%%%%%%
    next, we shall apply the Dolan and McCrea variational method\cite{Wanas2009} received via personal communication to each term of Eq.\eqref{eqn6-1::5d}. Before that, let's note the following; they would come in handy.
    
   \begin{equation}
   \begin{split}
      \frac{\partial\apvec{r}{\chi}_\alpha}{\partial\apvec{i}{\chi}_\rho}=\delta_{ir}\delta_\alpha^\rho;\qquad \frac{\partial\apvec{r}{\chi}^\alpha}{\partial\apvec{i}{\chi}_\rho}=-\apvec{r}{\chi}^\rho\apvec{i}{\chi}^\alpha=-\delta_{ir}\apvec{r}{\chi}^\rho\apvec{r}{\chi}^\alpha;\\\frac{\partial\apvec{r}{\chi}_{\alpha,\sigma}}{\partial\apvec{i}{\chi}_{\rho,\lambda}}=\delta_{ir}\delta_\alpha^\rho\delta_\sigma^\lambda;\qquad\frac{\partial\apvec{r}{\chi}^\alpha{}_{,}{}^\sigma}{\partial\apvec{i}{\chi}_{\rho,\lambda}}=\delta_{ir}g^{\alpha\rho}g^{\sigma\lambda}\label{e11}.                
    \end{split}
    \end{equation}
    Using Eq.\eqref{e11}, we show the derivative with respect to the parallelization vector of the metric tensor, torsion and contortion below.
    \begin{equation}
    \begin{split}
    \frac{\partial g^{\alpha\sigma}}{\partial\apvec{i}{\chi}_\rho}&=\frac{\partial}{\partial\apvec{i}{\chi}_\rho}(\apvec{r}{\chi}^\alpha\apvec{r}{\chi}^\sigma)\\
    &=\frac{\partial\apvec{r}{\chi}^\alpha}{\partial\apvec{i}{\chi}_\rho}\apvec{r}{\chi}^\sigma+\apvec{r}{\chi}^\alpha\frac{\partial\apvec{r}{\chi}^\sigma}{\partial\apvec{i}{\chi}_\rho}\\
    &=-\apvec{r}{\chi}^\rho\apvec{i}{\chi}^\alpha\apvec{r}{\chi}^\sigma-\apvec{r}{\chi}^\alpha\apvec{r}{\chi}^\rho\apvec{i}{\chi}^\sigma\\
    &=-\apvec{i}{\chi}^\alpha g^{\sigma\rho}-\apvec{i}{\chi}^\sigma g^{\alpha\rho}\label{eqn6-1::5e},
    \end{split}
    \end{equation} 
    \begin{equation}
    \begin{split}
    \frac{\partial\Lambda^\epsilon{}_{\sigma\lambda}}{\partial\apvec{i}{\chi}_\rho}&=\frac{\partial}{\partial\apvec{i}{\chi}_\rho}(\apvec{r}{\chi}^\epsilon\apvec{r}{\chi}_{\sigma,\lambda}-\apvec{r}{\chi}^\epsilon\apvec{r}{\chi}_{\lambda,\sigma})\\
    &=-\apvec{r}{\chi}^\rho\apvec{i}{\chi}^\epsilon\apvec{r}{\chi}_{\sigma,\lambda}+\apvec{r}{\chi}^\rho\apvec{i}{\chi}^\epsilon\apvec{r}{\chi}_{\lambda,\sigma}\\
    &=-\apvec{i}{\chi}^\epsilon(\apvec{r}{\chi}^\rho\apvec{r}{\chi}_{\sigma,\lambda}-\apvec{r}{\chi}^\rho\apvec{r}{\chi}_{\lambda,\sigma})\\
    &=-\apvec{i}{\chi}^\epsilon\Lambda^\rho{}_{\sigma\lambda}\label{eqn6-1::5f}.
    \end{split}
    \end{equation}
    %\textbf{First Term of \eqref{eqn6-1::5d}}\\[-0.1cm]\rule{\textwidth}{0.01cm}
    The derivative of the first term of Eq.\eqref{eqn6-1::5d} is given as:
    \begin{equation}
    \begin{split}
    \frac{\partial \chi}{\partial\apvec{i}{\chi}_\rho}=\chi\apvec{i}{\chi}^\rho\label{eqn6-1::5g}.
    \end{split}
    \end{equation}
    
    %\textbf{Second Term of \eqref{eqn6-1::5d}}\\[-0.5cm]\rule{\textwidth}{0.05cm}
    \begin{widetext}
    The derivative of the second term of Eq.\eqref{eqn6-1::5d} is given as:
    \begin{equation}
    \begin{split}
    \frac{\partial\Lambda^{\epsilon\alpha\mu}}{\partial\apvec{i}{\chi}_\rho}
    &=\frac{\partial}{\partial\apvec{i}{\chi}_\rho}(g^{\alpha\sigma}g^{\mu\lambda}\Lambda^\epsilon{}_{\sigma\lambda})\\
    &=\frac{\partial g^{\alpha\sigma}}{\partial\apvec{i}{\chi}_\rho}g^{\mu\lambda}\Lambda^\epsilon{}_{\sigma\lambda}+g^{\alpha\sigma}\frac{\partial g^{\mu\lambda}}{\partial\apvec{i}{\chi}_\rho}\Lambda^\epsilon{}_{\sigma\lambda}+g^{\alpha\sigma}g^{\mu\lambda}\frac{\partial\Lambda^\epsilon{}_{\sigma\lambda}}{\partial\apvec{i}{\chi}_\rho}\\
    &=(-\apvec{i}{\chi}^\alpha g^{\sigma\rho}-\apvec{i}{\chi}^\sigma g^{\alpha\rho})g^{\mu\lambda}\Lambda^\epsilon{}_{\sigma\lambda}+g^{\alpha\sigma}(-\apvec{i}{\chi}^\mu g^{\lambda\rho}-\apvec{i}{\chi}^\lambda g^{\mu\rho})\Lambda^\epsilon{}_{\sigma\lambda}\\&\qquad-\apvec{i}{\chi}^\epsilon g^{\alpha\sigma}g^{\mu\lambda}\Lambda^\rho{}_{\sigma\lambda}\\	
    &=-\apvec{i}{\chi}^\alpha g^{\sigma\rho}g^{\mu\lambda}\Lambda^\epsilon{}_{\sigma\lambda}-\apvec{i}{\chi}^\sigma g^{\alpha\rho}g^{\mu\lambda}\Lambda^\epsilon{}_{\sigma\lambda}-\apvec{i}{\chi}^\lambda g^{\alpha\sigma}g^{\mu\rho}\Lambda^\epsilon{}_{\sigma\lambda}\\&\qquad-\apvec{i}{\chi}^\mu g^{\alpha\sigma}g^{\lambda\rho}\Lambda^\epsilon{}_{\sigma\lambda}-\apvec{i}{\chi}^\epsilon g^{\alpha\sigma}g^{\mu\lambda}\Lambda^\rho{}_{\sigma\lambda}\\
    &=-\apvec{i}{\chi}^\alpha\Lambda^{\epsilon\rho\mu}-\apvec{i}{\chi}_\gamma g^{\gamma\sigma}g^{\alpha\rho}g^{\mu\lambda}\Lambda^\epsilon{}_{\sigma\lambda}-\apvec{i}{\chi}_\gamma g^{\gamma\lambda}g^{\alpha\sigma}g^{\mu\rho}\Lambda^\epsilon{}_{\sigma\lambda}\\&\qquad-\apvec{i}{\chi}^\mu\Lambda^{\epsilon\alpha\rho}-\apvec{i}{\chi}^\epsilon\Lambda^{\rho\alpha\mu}\\
    &=-\apvec{i}{\chi}^\alpha\Lambda^{\epsilon\rho\mu}-\apvec{i}{\chi}_\gamma(g^{\alpha\rho}\Lambda^{\epsilon\gamma\mu}+g^{\mu\rho}\Lambda^{\epsilon\alpha\gamma})-\apvec{i}{\chi}^\mu\Lambda^{\epsilon\alpha\rho}-\apvec{i}{\chi}^\epsilon\Lambda^{\rho\alpha\mu}\label{eqn6-1::5h}.
    \end{split}
    \end{equation}
\end{widetext}

\begin{widetext}
The derivative of the third term from Eq.\eqref{eqn6-1::5d} is given as:
   \begin{equation}
    \begin{split}
    \frac{\partial\gamma_{\alpha\mu\epsilon}}{\partial\apvec{i}{\chi}_\rho}
    &=\frac{1}{2}\frac{\partial}{\partial\apvec{i}{\chi}_\rho}\Big[\Lambda_{\mu\epsilon\alpha}+\Lambda_{\epsilon\mu\alpha}-\Lambda_{\alpha\epsilon\mu} \Big]\\
    &=\frac{1}{2}\frac{\partial}{\partial\apvec{i}{\chi}_\rho}\Big[\apvec{r}{\chi}_\mu(\apvec{r}{\chi}_{\epsilon,\alpha}-\apvec{r}{\chi}_{\alpha,\epsilon})+\apvec{r}{\chi}_\epsilon(\apvec{r}{\chi}_{\mu,\alpha}-\apvec{r}{\chi}_{\alpha,\mu})-\apvec{r}{\chi}_\alpha(\apvec{r}{\chi}_{\epsilon,\mu}-\apvec{r}{\chi}_{\mu,\epsilon})\Big]\\
    &=\frac{1}{2}\delta_{ir}\Big[\delta^\rho_\mu(\apvec{r}{\chi}_{\epsilon,\alpha}-\apvec{r}{\chi}_{\alpha,\epsilon})+\delta^\rho_\epsilon(\apvec{r}{\chi}_{\mu,\alpha}-\apvec{r}{\chi}_{\alpha,\mu})-\delta^\rho_\alpha(\apvec{r}{\chi}_{\epsilon,\mu}-\apvec{r}{\chi}_{\mu,\epsilon})\Big]\\
    &=\frac{1}{2}\apvec{i}{\chi}_\lambda\Big[\delta^\rho_\mu\apvec{r}{\chi}^\lambda(\apvec{r}{\chi}_{\epsilon,\alpha}-\apvec{r}{\chi}_{\alpha,\epsilon})+\delta^\rho_\epsilon\apvec{r}{\chi}^\lambda(\apvec{r}{\chi}_{\mu,\alpha}-\apvec{r}{\chi}_{\alpha,\mu})-\delta^\rho_\alpha\apvec{r}{\chi}^\lambda(\apvec{r}{\chi}_{\epsilon,\mu}-\apvec{r}{\chi}_{\mu,\epsilon})\Big]\\
    &=\frac{1}{2}\apvec{i}{\chi}_\lambda\Big[\delta^\rho_\mu\Lambda^\lambda{}_{\epsilon\alpha}+\delta^\rho_\epsilon\Lambda^\lambda{}_{\mu\alpha}-\delta^\rho_\alpha\Lambda^\lambda{}_{\epsilon\mu}\Big]
    \label{eqn6-1::5i}.
    \end{split}
    \end{equation}
\end{widetext}
\begin{widetext}
Now, we put together all the terms, that is, the first, second and third terms of Eq.\eqref{eqn6-1::5d} to obtain:
    \begin{equation}
    \begin{split}
    {1\over\chi}{\partial Q\over\partial\apvec{i}{\chi}_\rho}
    &={1\over2}q^2\big[\apvec{i}{\chi}^\rho\Lambda^{\epsilon\alpha\mu}-\apvec{i}{\chi}^\alpha\Lambda^{\epsilon\rho\mu}-\apvec{i}{\chi}^\mu\Lambda^{\epsilon\alpha\rho}-\apvec{i}{\chi}^\epsilon\Lambda^{\rho\alpha\mu}-\apvec{i}{\chi}_\gamma\big( g^{\alpha\rho}\Lambda^{\epsilon\gamma\mu}+g^{\mu\rho}\Lambda^{\epsilon\alpha\gamma} \big)\big]\gamma_{\alpha\mu\epsilon}\\
    &\qquad+{1\over4}q^2\apvec{i}{\chi}_\lambda\Lambda^{\epsilon\alpha\mu}\big( \delta^\rho_\mu\Lambda^\lambda{}_{\epsilon\alpha}+\delta^\rho_\epsilon\Lambda^\lambda{}_{\mu\alpha}-\delta^\rho_\alpha\Lambda^\lambda{}_{\epsilon\mu} \big)\label{eqn6-1::5j}.
    \end{split}
    \end{equation}
\end{widetext}
\begin{widetext}
    Next, we compute the second term in the Euler-Lagrange equation stated at Eq.\eqref{eqn6-1::5c}, that is:
    \begin{equation}
    \begin{split}
    \frac{\partial Q}{\partial\apvec{i}{\chi}_{\rho,\eta}}
    &=\frac{\partial}{\partial\apvec{i}{\chi}_{\rho,\eta}}\Big(\frac{1}{2}q^2\chi\Lambda^{\epsilon\alpha\mu}\gamma_{\alpha\mu\epsilon}\Big)\\
    &={1\over2}q^2\chi{\partial\Lambda^{\epsilon\alpha\mu}\over\partial\apvec{i}{\chi}_{\rho,\eta}}\gamma_{\alpha\mu\epsilon}+{1\over2}q^2\chi\Lambda^{\epsilon\alpha\mu}{\partial\gamma_{\alpha\mu\epsilon}\over\partial\apvec{i}{\chi}_{\rho,\eta}}\label{eqn6-1::5k}.
    \end{split}
    \end{equation}
\end{widetext}    
    %\textbf{From first term of \eqref{eqn6-1::5k}:}\\[-0.5cm]\rule{\textwidth}{0.05cm}
\begin{widetext}
    From the first term of Eq.\eqref{eqn6-1::5k}, we have:
    \begin{equation}
    \begin{split}
    {\partial\Lambda^{\epsilon\alpha\mu}\over\partial\apvec{i}{\chi}_{\rho,\eta}}
    &={\partial\over\partial\apvec{i}{\chi}_{\rho,\eta}}\Big[\apvec{r}{\chi}^\epsilon\big(\apvec{r}{\chi}^\alpha{}_,{}^\mu-\apvec{r}{\chi}^\mu{}_,{}^\alpha\big)\Big]\\
    &=\apvec{r}{\chi}^\epsilon\delta_{ir}\Big[g^{\alpha\rho}g^{\mu\eta}-g^{\mu\rho}g^{\alpha\eta}\Big]\\
    &=\apvec{i}{\chi}^\epsilon\Big[g^{\alpha\rho}g^{\mu\eta}-g^{\mu\rho}g^{\alpha\eta}\Big]\label{eqn6-1::5l}.
    \end{split}
    \end{equation}
\end{widetext}
    %\textbf{From second term of \eqref{eqn6-1::5k}:}\\[-0.5cm]\rule{\textwidth}{0.05cm}
\begin{widetext}   
    And from the second term of Eq.\eqref{eqn6-1::5k}, we have:
    \begin{equation}
    \begin{split}
    {\partial\gamma_{\alpha\mu\epsilon}\over\partial\apvec{i}{\chi}_{\rho,\eta}}
    &={\partial\over\partial\apvec{i}{\chi}_{\rho,\eta}}\Big[{1\over2}\big(\Lambda_{\mu\epsilon\alpha}+\Lambda_{\epsilon\mu\alpha}-\Lambda_{\alpha\epsilon\mu}\big)\Big]\\
    &={1\over2}{\partial\over\partial\apvec{i}{\chi}_{\rho,\eta}}\Big[\apvec{r}{\chi}_\mu(\apvec{r}{\chi}_{\epsilon,\alpha}-\apvec{r}{\chi}_{\alpha,\epsilon})  + \apvec{r}{\chi}_\epsilon(\apvec{r}{\chi}_{\mu,\alpha}-\apvec{r}{\chi}_{\alpha,\mu})  -\apvec{r}{\chi}_\alpha(\apvec{r}{\chi}_{\epsilon,\mu}-\apvec{r}{\chi}_{\mu,\epsilon})\Big]\\
    &={1\over2}\Big[\apvec{r}{\chi}_\mu\delta_{ir}(\delta^\rho_\epsilon\delta^\eta_\alpha-\delta^\rho_\alpha\delta^\eta_\epsilon)  +  \apvec{r}{\chi}_\epsilon\delta_{ir}(\delta^\rho_\mu\delta^\eta_\alpha-\delta^\rho_\alpha\delta^\eta_\mu) - \apvec{r}{\chi}_\alpha\delta_{ir}(\delta^\rho_\epsilon\delta^\eta_\mu-\delta^\rho_\mu\delta^\eta_\epsilon)\Big]\\
    &={1\over2}\Big[\apvec{i}{\chi}_\mu(\delta^\rho_\epsilon\delta^\eta_\alpha-\delta^\rho_\alpha\delta^\eta_\epsilon)  +  \apvec{i}{\chi}_\epsilon(\delta^\rho_\mu\delta^\eta_\alpha-\delta^\rho_\alpha\delta^\eta_\mu) - \apvec{i}{\chi}_\alpha(\delta^\rho_\epsilon\delta^\eta_\mu-\delta^\rho_\mu\delta^\eta_\epsilon)\Big]\label{eqn6-1::5m}.
    \end{split}
    \end{equation}
\end{widetext}
\begin{widetext}
    Putting Eq.\eqref{eqn6-1::5l} and Eq.\eqref{eqn6-1::5m} into Eq.\eqref{eqn6-1::5k}, we have:
    \begin{equation}
    \begin{split}
    {\partial Q\over\partial\apvec{i}{\chi}_{\rho,\eta}}={1\over2}q^2\chi\apvec{i}{\chi}^\epsilon(g^{\alpha\rho}g^{\mu\eta}-g^{\mu\rho}g^{\alpha\eta})\gamma_{\alpha\mu\epsilon}+{1\over4}q^2\chi\Lambda^{\epsilon\alpha\mu}\Big[\apvec{i}{\chi}_\mu\delta^{\rho\eta}_{[\epsilon\alpha]} + \apvec{i}{\chi}_\epsilon\delta^{\rho\eta}_{[\mu\alpha]} + \apvec{i}{\chi}\delta^{\rho\eta}_{[\mu\epsilon]}  \Big]
    \end{split}\label{eqn6-1::5n}.
    \end{equation}
\end{widetext}
\begin{widetext}
    Next, we carry out a derivative of Eq.\eqref{eqn6-1::5n} with respect to $x^\eta$ to obtain:\footnote{It can be shown that $\chi_{,\eta}=\chi\apvec{j}{\chi}^\sigma\apvec{j}{\chi}_{\sigma,\eta}$}
    \begin{equation}
    \begin{split}
    {1\over\chi}\Bigg( {\partial Q\over\partial\apvec{i}{\chi}_{\rho,\eta} } \Bigg)_{,\eta}=&\quad{1\over2}q^2\big(g^{\alpha\rho}g^{\mu\eta}-g^{\mu\rho}g^{\alpha\eta}\big)\big(\Gamma^\sigma_{\sigma\eta}\apvec{i}{\chi}^\epsilon\gamma_{\alpha\mu\epsilon}+\apvec{i}{\chi}^\epsilon_{,\eta}\gamma_{\alpha\mu\epsilon}+\apvec{i}{\chi}^\epsilon\gamma_{\alpha\mu\epsilon,\eta}\big)\\
    &+{1\over2}q^2\apvec{i}{\chi}^\epsilon\big(g^{\alpha\rho}{}_{,\eta}g^{\mu\eta}+g^{\alpha\rho}g^{\mu\eta}{}_{,\eta}-g^{\mu\rho}{}_{,\eta}g^{\alpha\eta}-g^{\mu\rho}g^{\alpha\eta}{}_{,\eta}\big)\gamma_{\alpha\mu\epsilon}\\
    &+{1\over4}q^2\big(\Gamma^\sigma_{\sigma\eta}\Lambda^{\epsilon\alpha\mu}+\Lambda^{\epsilon\alpha\mu}{}_{,\eta}\big)\big(\apvec{i}{\chi}_\mu\delta^{\rho\eta}_{[\epsilon\alpha]}+\apvec{i}{\chi}_\epsilon\delta^{\rho\eta}_{[\mu\alpha]}+\apvec{i}{\chi}_\alpha\delta^{\rho\eta}_{[\mu\epsilon]}\big)\\
    &+{1\over4}q^2\Lambda^{\epsilon\alpha\mu}\big(\apvec{i}{\chi}_{\mu,\eta}\delta^{\rho\eta}_{[\epsilon\alpha]}+\apvec{i}{\chi}_{\epsilon,\eta}\delta^{\rho\eta}_{[\mu\alpha]}+\apvec{i}{\chi}_{\alpha,\eta}\delta^{\rho\eta}_{[\mu\epsilon]}\big)\label{eqn6-1::5o}.
    \end{split}
    \end{equation}
\end{widetext}
\begin{widetext}
    Putting the result of Eq.\eqref{eqn6-1::5j} and Eq.\eqref{eqn6-1::5o} into Eq.\eqref{eqn6-1::5c} and simplifying, we get\footnote{$\gamma^{\alpha\rho}{}_{\nu||\alpha}$: the vertical bars denote covariant differenttiation with respect to the parameterized canonical AP-connection.}:
    \begin{equation}
    \begin{split}
    0&=2G^\rho_\nu-q^2\delta^\rho_\nu\gamma^{\epsilon\alpha\mu}\gamma_{\alpha\mu\epsilon}+2q^2\gamma^{\alpha\rho}{}_{\epsilon}\gamma^\epsilon{}_{\alpha\nu}\\
    &+2q^3c_\alpha\gamma^{\alpha\rho}{}_{\nu}-2q^3\gamma^{\alpha\rho}{}_{\epsilon}\gamma^\epsilon{}_{\nu\alpha}-2q^3\gamma^{\alpha\rho}{}_{\epsilon}\gamma^\epsilon{}_{\alpha\nu}-2q^2\gamma^{\alpha\rho}{}_{\nu||\alpha}\label{eqn6-1::5p}.
    \end{split}
    \end{equation}
    The result of Eq.\eqref{eqn6-1::5p} agrees with \cite{WanasEtal2015}. Hence, one of our goals, which is to reinvent the undocumented Dolan-McCrea variational method is successful.
\end{widetext}
\begin{widetext}
    It's easy to see that multiplying Eq.\eqref{eqn6-1::5p} through by $g_{\lambda\rho}$, gives%%\footnote{$\gamma^\alpha{}_{\lambda\nu||\alpha}$ stands for covariant differentiation with respect to the parameterized canonical connection.}:
    
    \begin{equation}
    \begin{split}
    0&=2G_{\nu\lambda}-q^2g_{\nu\lambda}\gamma^{\epsilon\alpha\mu}\gamma_{\alpha\mu\epsilon}+2q^2\gamma^\alpha{}_{\lambda\epsilon}\gamma^\epsilon{}_{\alpha\nu}\\
    &+2q^3c_\alpha\gamma^\alpha{}_{\lambda\nu}-2q^3\gamma^\alpha{}_{\lambda\epsilon}\gamma^\epsilon{}_{\nu\alpha}-2q^3\gamma^\alpha{}_{\lambda\epsilon}\gamma^\epsilon{}_{\alpha\nu}-2q^2\gamma^\alpha{}_{\lambda\nu||\alpha}\label{eqn6-1::5q}.		
    \end{split}
    \end{equation}
\end{widetext} 
    Since the above entirely geometric object is non-symmetric, it can be decomposed into symmetric and skew-symmetric parts. The symmetric part will be seen to comprise the Einstein tensor $G_{\nu\lambda}$ plus other symmetric terms, which have been collectively identified as the geometric energy-momentum tensor. The Einstein tensor is a geometric object, the energy-momentum tensor of Einstein GR is a phenomenological object but in this situation, the energy-momentum tensor is a geometric consequence of the AP-geometry\cite{WanasEtal2015}.
    
    %\subsection{Second Order World Tensors}
    M.I. Wanas and others have defined second order world tensors in the modified AP-space\cite{WanasOsmanKholy2015}. Before we proceed in decomposing the non-symmetric field equations at  Eq.\eqref{eqn6-1::5q} into symmetric and skew-symmetric components, it is pertinent at this juncture to introduce the relevant second order world tensors so we can keep the math less messy to the sight.
    \begin{table}[h!]
        \centering
        \caption{Second Order Symmetric World Tensors\footnote{{\fontsize{10}{15}\selectfont $\nabla^\varepsilon{}_{\varsigma\alpha}$ is the parameterized canonical connection. Every term in the table is parameterized; the parameters are suppressed for brevity.}}\cite{WanasOsmanKholy2015}\label{tab6-1::1}}
        %\tabulinesep=0.8mm
        %{
            \begin{tabular}{ll}
                S/N& World tensors\\
                %\rowcolor{black!30}
                1.\label{1} & $\psi_{\nu\lambda}:=\triangle^\alpha{}_{\nu\lambda||\alpha}$\\
                2.\label{2} & $\phi_{\nu\lambda}:=C_\alpha\triangle^\alpha{}_{\nu\lambda}$\\
                %\rowcolor{black!30}
                3.\label{3} & $\omega_{\nu\lambda}:=\gamma^\alpha{}_{\nu\epsilon}\gamma^\epsilon{}_{\lambda\alpha}$\\
                4.\label{4} & $\varpi_{\nu\lambda}:=\gamma^\alpha{}_{\nu\epsilon}\gamma^\epsilon{}_{\alpha\lambda}+\gamma^\alpha{}_{\lambda\epsilon}\gamma^\epsilon{}_{\alpha\nu}$\\
            \end{tabular}    
        %}
    \end{table}
    
    Let's for mathematical convenience, denote Eq.\eqref{eqn6-1::5q} simply by $B_{\nu\lambda}$ and now we decompose it as follows:
    \begin{equation}
    \begin{split}
    0&=B_{\nu\lambda}\\
    &=2G_{\nu\lambda}-q^2g_{\nu\lambda}\gamma^{\epsilon\alpha\mu}\gamma_{\alpha\mu\epsilon}+2q^2\gamma^\alpha{}_{\lambda\epsilon}\gamma^\epsilon{}_{\alpha\nu}+2q^3c_\alpha\gamma^\alpha{}_{\lambda\nu}\\&\quad-2q^3\gamma^\alpha{}_{\lambda\epsilon}\gamma^\epsilon{}_{\nu\alpha}-2q^3\gamma^\alpha{}_{\lambda\epsilon}\gamma^\epsilon{}_{\alpha\nu}-2q^2\gamma^\alpha{}_{\lambda\nu||\alpha}\label{eqn6-1::5r}\\
    &={1\over2}\big(B_{\nu\lambda} + B_{\lambda\nu}\big)+{1\over2}\big(B_{\nu\lambda} - B_{\lambda\nu}\big)\\
    &=B_{(\nu\lambda)} + B_{[\nu\lambda]}.
    \end{split}
    \end{equation}
    Since each $B_{\nu\lambda}$ in the decomposition equals zero, both the symmetric and the skew-symmetric components must each equals zero. The symmetric part $B_{(\nu\lambda)}$ is extracted and displayed as:
    \begin{equation}
    \begin{split}
    0&=B_{(\nu\lambda)}\\
    &=\big(G_{\nu\lambda}+G_{\lambda\nu}\big)-{1\over2}q^2\big(g_{\nu\lambda}+g_{\lambda\nu}\big)\gamma^{\epsilon\alpha\mu}\gamma_{\alpha\mu\epsilon}\\
    &+q^2\big(\gamma^\alpha{}_{\lambda\epsilon}\gamma^\epsilon{}_{\alpha\nu}+\gamma^\alpha{}_{\nu\epsilon}\gamma^\epsilon{}_{\alpha\lambda}\big)+q^3C_\alpha\big(\gamma^\alpha{}_{\lambda\nu}+\gamma^\alpha{}_{\nu\lambda}\big)\\&-q^3\big(\gamma^\alpha{}_{\lambda\epsilon}\gamma^\epsilon{}_{\nu\alpha}+\gamma^\alpha{}_{\nu\epsilon}\gamma^\epsilon{}_{\lambda\alpha}\big)
    -q^3\big(\gamma^\alpha{}_{\lambda\epsilon}\gamma^\epsilon{}_{\alpha\nu}+\gamma^\alpha{}_{\nu\epsilon}\gamma^\epsilon{}_{\alpha\lambda}\big)\\&-q^3\big(\gamma^\alpha{}_{\lambda\nu||\alpha}+\gamma^\alpha{}_{\nu\lambda||\alpha}\big)\label{eqn6-1::5s}.
    \end{split}
    \end{equation}
    Using the information in Table \ref{tab6-1::1} and noting that $\mathbf{\gamma.\gamma}+\mathbf{\gamma.\gamma}=g^{\nu\lambda}\varpi_{\nu\lambda}=\varpi$, we write Eq.\eqref{eqn6-1::5s} as:
    \begin{equation}
    \begin{split}
    0=&2G_{\nu\lambda}-{1\over2}q^2g_{\nu\lambda}\varpi+q^2\varpi_{\nu\lambda}+q^3\phi_{\nu\lambda}\\&\quad-2q^3\omega_{\nu\lambda}-q^3\varpi_{\nu\lambda}-q^3\psi_{\nu\lambda}\label{eqn6-1::5t}.	
    \end{split}
    \end{equation} 
    The original Einstein equation of gravity has the geometric Einstein tensor $G_{\nu\lambda}$, equals a phenomenological energy-mementum tensor $T_{\nu\lambda}$. We now write Eq.\eqref{eqn6-1::5t} to reflect that structural appearance as: 
    \begin{equation}
    \begin{split}
    G_{\nu\lambda}=&-{1\over2}\Big(-{1\over2}q^2g_{\nu\lambda}\varpi+q^2\varpi_{\nu\lambda}+q^3\phi_{\nu\lambda}\\&-2q^3\omega_{\nu\lambda}-q^3\varpi_{\nu\lambda}-q^3\psi_{\nu\lambda}\Big)\label{eqn6-1::5u},
    \end{split}
    \end{equation}
    %%%%The term in brackets in Eq.\eqref{eqn6-1::5u} has been likened to the energy-momentum tensor, $T_{\nu\lambda}$ of the conventional %%%%%Einstein's gravity theory. This term \footnote{The under-set letter `g' at $\apvec{g}{T}_{\nu\lambda}$ only serves to differentiate %%%%the %geometric energy-momentum distribution from the phenomenological energy-momentum distribution $T_{\nu\lambda}$ of Einstein GR.} %%%%$\apvec{g}{T}_{\nu\lambda}$, is a pure geometric object, it emerges naturally from the geometry just like the Einstein tensor %%%%$G_{\nu\lambda}$, whereas in the Einstein gravity theory, the Einstein tensor is geometrical and the energy-momentum tensor %%%%$T_{\nu\lambda}$ is phenomenological. 
    with:
    \begin{equation}
    \begin{split}
    \apvec{g}{T}_{\nu\lambda}=&q^2\big(\varpi_{\nu\lambda}-{1\over2}g_{\nu\lambda}\varpi\big)\\&+q^3\big(\phi_{\nu\lambda}-2\omega_{\nu\lambda}-\varpi_{\nu\lambda}-\psi_{\nu\lambda}\big)\label{eqn6-1::5v}
    \end{split}
    \end{equation}
    The arrow at Eq.\eqref{cor} implies a correspondence between the geometric tensor $\apvec{g}{T}_{\nu\lambda}$ with the phenomenological energy-momentum tensor $T_{\nu\lambda}$. 
    \begin{equation}
        \begin{split}
          \apvec{g}{T}_{\nu\lambda}\rightarrow T_{\nu\lambda}\label{cor}.
        \end{split}
    \end{equation}
    
    In a more compact and yet more informative form, we write Eq.\eqref{eqn6-1::5u} as:
    \begin{equation}
    \begin{split}
    R_{\nu\lambda}-{1\over2}g_{\nu\lambda}R=-{1\over2}\apvec{g}{T}_{\nu\lambda}\label{eqn6-1::5w}.
    \end{split}
    \end{equation}
    
    It easy to see that setting $q=0$, reduces Eq.\eqref{eqn6-1::5w} to the case of Einstein equation in free space. This implies Riemannian geometry hides away this aspect of nature, hence, Einstein obtained gravity equations that are partly geometric and partly phenomenological. This suggests that the present geometry is superior to the Riemannian in unraveling the intricacies of nature.
    
    In what follows, we write the skew-symmetric part $B_{[\nu\lambda]}$ of Eq.\eqref{eqn6-1::5q}. Before that, we also tabulate the relevant second order world skew-symmetric tensors. As we said earlier, this helps to keep the math less fuzzy.
    \begin{table}[h!]
        \centering
        \caption{Second Order skew-symmetric World tensors\footnote{{\fontsize{10}{15}\selectfont Every term in the table is parameterized; the parameters are suppressed for brevity.}}\cite{WanasOsmanKholy2015}\label{tab6-1::2}}
        %\tabulinesep=0.8mm
        %{
            \begin{tabular}{ll}
                S/N				&	World tensors\\
                1.\label{1.}	& $\kappa_{\nu\lambda}:=\gamma^\alpha{}_{\nu\epsilon}\gamma^\epsilon{}_{\alpha\lambda}-\gamma^\alpha{}_{\lambda\epsilon}\gamma^\epsilon{}_{\alpha\nu}$	\\
                2.\label{2.}	&	$\chi_{\nu\lambda}:=\Lambda^\alpha{}_{\nu\lambda||\alpha}$ \\
                3.\label{3.}	&	$\eta_{\nu\lambda}:=C_\alpha\Lambda^\alpha{}_{\nu\lambda}$\\
                4.\label{4.}	&	$\epsilon_{\nu\lambda}:=C_{\nu||\lambda}-C_{\lambda||\nu}$\\
            \end{tabular}            
           \end{table}
    
    Also, let us note the following important identity satisfied by skew-symmetric tensors; it will come in handy.
    \begin{equation}
    \begin{split}
    \eta_{\nu\lambda}+\epsilon_{\nu\lambda}-\chi_{\nu\lambda}\equiv0\label{eqn6-1::5x}.
    \end{split}
    \end{equation} 
    Let us also elaborate mathematically on item 4 of Table \ref{tab6-1::2}.
    \begin{equation}
    \begin{split}
    \epsilon_{\nu\lambda}&:=C_{\nu||\lambda}-C_{\lambda||\nu}\\
    &=\big(C_{\nu,\lambda}-C_{\lambda,\nu}\big)+C_\alpha\Big(\{^\alpha_{\lambda\nu}\}-\gamma^\alpha{}_{\lambda\nu}-\{^\alpha_{\nu\lambda}\}+\gamma^\alpha{}_{\nu\lambda}\Big)\\
    &=\big(C_{\nu,\lambda}-C_{\lambda,\nu}\big)-C_\alpha\Lambda^\alpha{}_{\nu\lambda}\\
    &=\big(C_{\nu,\lambda}-C_{\lambda,\nu}\big)-\eta_{\nu\lambda}\label{eqn6-1::5y}.
    \end{split}
    \end{equation}
    
    Now, using Eq.\eqref{eqn6-1::5y} and Table \ref{tab6-1::2}, the skew-symmetric part of Eq.\eqref{eqn6-1::5q} is: 
    \begin{equation}
    \begin{split}
    0	&= B_{[\nu\lambda]}\\
    &=q^2\big(\gamma^\alpha{}_{\lambda\epsilon}\gamma^\epsilon{}_{\alpha\nu}-\gamma^\alpha{}_{\nu\epsilon}\gamma^\epsilon{}_{\alpha\lambda}\big)+q^3C_\alpha\big( \gamma^\alpha{}_{\lambda\nu}-\gamma^\alpha{}_{\nu\lambda}\big)\\
    &\quad-q^3\big(\gamma^\alpha{}_{\lambda\epsilon}\gamma^\epsilon{}_{\alpha\nu}-\gamma^\alpha{}_{\nu\epsilon}\gamma^\epsilon{}_{\alpha\lambda}\big)-q^3\big(\gamma^\alpha{}_{\lambda\nu}-\gamma^\alpha{}_{\nu\lambda}\big)_{||\alpha}\\
    &=-\kappa_{\nu\lambda}+q\kappa_{\nu\lambda}+q\big(\chi_{\nu\lambda}-\eta_{\nu\lambda}\big)\\
    &=-(1-q)\kappa_{\nu\lambda}+q\epsilon_{\nu\lambda}\\
    &=-(1-q)\kappa_{\nu\lambda}+q\big( -\eta_{\nu\lambda}+C_{\nu,\lambda}-C_{\lambda,\nu} \big)\label{eqn6-1::5z}.
    \end{split}
    \end{equation}
    Recognizing $(1-q)\kappa_{\nu\lambda}+q\eta_{\nu\lambda}=F_{\nu\lambda}$, we write:
    \begin{equation}
    \begin{split}
    F_{\nu\lambda}=q\big(C_{\nu,\lambda}-C_{\lambda,\nu}\big)\label{eqn6.6a}.
    \end{split}
    \end{equation}
    %We can see that there is logical correspondence between the present combined field equations with the nonlinear field equations. 
    
    Eq.\eqref{eqn6-1::5u} shows a very agreeable morphological correspondence with the nonlinear equations of general relativity due to Einstein. Eq.\eqref{eqn6.6a} resembles very much the electrognetic field tensor due to Maxwell. The difference between the present theory and the nonlinear theories due to Einstein and due to Maxwell is that the present theory is totally geometrical. Einstein general relativity is partly geometrical and partly phenomenological. In what follows, we apply a linearization scheme on the symmetric and skew-symmetric equations so we can make comparison with linear field theories.
    \section{Linearized Field Equations}
    %In the previous section, the symmetric part of the combined field equations has been compared and found to correspond with the Einstein %general relativity. Same was done for the skew-symmetric part; correspondence was achieved with the Einstein-Maxwell theory. But these %equations, with which the present theory is compared, are nonlinear.
    
    For Eq.\eqref{eqn6-1::5w} and Eq.\eqref{eqn6.6a} to be physically and experimentally viable, the equations should have proper classical behavior. So, we shall linearize them and compare the results with linear theories notably those of Newton. The linearization method to be used was been put forward by F.I. Mikhail and M.I. Wanas\cite{WanasMikhailGFT1981}. This method involves a series expansion of the tetrad or parallelization vector of AP-space. We shall present the rudiments of the arithmetic that this entails and we shall apply it in getting linearized expressions for regularly encounter AP geometric objects. We shall then re-substitute linearized versions of these objects into the field equations.
    
    On the parallelizable manifold, the Lorentzian metric $g_{\mu\nu}$ comprises product of two tetrads. The tetrad is in general spacetime dependent. It would be spacetime independent if we intend to build the Minkowski metric from a tetrad, in which case we have the Minkowski metric as:
    \begin{equation}
    \begin{split}
    \eta^\prime_{\mu\nu}:=\apvec{i}{\chi^\prime}_\mu\apvec{i}{\chi^\prime}_\nu\label{eqnLin:a}.
    \end{split}
    \end{equation}
    The perturbative expansion of the spacetime dependent tetrad is given at Eq.\eqref{eqnLin:c}. The expansion parameter s is so small that we ignore squared and higher order terms; this implies we are considering weak field approximations in this scheme. 
    \begin{equation}
    \begin{split}
    \apvec{i}{\chi}_\mu(x):=\delta_{i\mu}+s\apvec{i}{U}_\mu(x)\label{eqnLin:c}.
    \end{split}
    \end{equation} 
    Expansion of the contravariant tetrad field is given below.
    \begin{equation}
    \begin{split}
    \apvec{i}{\chi}^\mu(x):=\delta_{i\mu}-s\apvec{\mu}{U}_i(x)+O(\lambda^2)\label{eqnLin:d}.
    \end{split}
    \end{equation}
    
    With the above definition for the tetrad field, every geometric object of the AP-space can be constructed in terms of it. For instance, we present in what follow, s-linearized expressions for the relevant AP objects.
    
    \begin{equation}
    \begin{split}
    g_{\nu\lambda} &=\apvec{i}{\chi}_\nu\apvec{i}{\chi}_\lambda\\
    &=(\delta_{i\nu}+s\apvec{i}{U}_\nu)(\delta_{i\lambda}+s\apvec{i}{U}_\lambda)\\
    &=\delta_{\nu\lambda}+s(\apvec{\nu}{U}_\lambda+\apvec{\lambda}{U}_\nu)+s^2\apvec{i}{U}_\nu\apvec{i}{U}_\lambda\\
    &=\delta_{\nu\lambda}+sW_{\nu\lambda}+s^2\apvec{i}{U}_\nu\apvec{i}{U}_\lambda\label{eqnLin:e}.
    \end{split}
    \end{equation}
    
    Similarly, we obtain for other geometric AP objects the following expressions\cite{NabilnOthers2013, WanasEtal2015}.
    \begin{equation}
    \begin{split}
    \{^\alpha_{\nu\lambda}\}={s\over2}(W_{\nu\alpha,\lambda}+W_{\lambda\alpha,\nu}-W_{\nu\lambda,\alpha})\label{eqnLin:f},
    \end{split}
    \end{equation}
    where $W_{\nu\lambda}=(\apvec{\lambda}{U}_\nu+\apvec{\nu}{U}_\lambda)$.
    
    \begin{equation}
    \begin{split}
    \Gamma^\alpha_{\mu\nu}=s\apvec{\alpha}{U}_{\mu,\nu}\label{eqnLin:g}
    \end{split}
    \end{equation}
    The parameterized torsion, contortion and basic form are given respectively as:
    \begin{equation}
    \begin{split}
    \Lambda^\alpha_{\mu\nu}=sq(\apvec{\alpha}{U}_{\mu,\nu}-\apvec{\alpha}{U}_{\nu,\mu})\label{eqnLin:h},
    \end{split}
    \end{equation}
    \begin{equation}
    \begin{split}
    \gamma^\alpha_{\mu\nu}=sq\apvec{\alpha}{U}_{\mu,\nu}-{s\over2}q(W_{\mu\alpha,\nu}+W_{\nu\alpha,\mu}-W_{\mu\nu,\alpha})\label{eqnLin:i},
    \end{split}
    \end{equation}
    \begin{equation}
    \begin{split}
    C_\mu=sq(\apvec{\alpha}{U}_{\mu,\alpha}-\apvec{\alpha}{U}_{\alpha,\mu})\label{eqnLin:j}.
    \end{split}
    \end{equation}
    Recall that q is the parameterization term, where s is the perturbation parameter. The following Table \ref{tab6-1::3} gives the least order and maximum order of the perturbation parameter of various linearized geometric objects of AP-space.
    
    \begin{table}[h!]
        \centering
        \caption{Order\footnote{\checkmark: implies 2 and higher powers of s} of Perturbation of AP tensors\cite{WanasOsmanKholy2015}\label{tab6-1::3}}
        \begin{tabular}{p{3cm}p{3cm}}
            Geometric object &  Power of s\\
        \end{tabular}
        \rule{0.4\textwidth}{0.2mm}
        \begin{tabular}{p{3cm}p{1.5cm}p{1.5cm}}
            &  Minimum & Maximum \\
            
            $\apvec{i}{\chi}_\nu$ 			 &  0  		& 1			\\
            \rowcolor{black!30}
            $g_{\nu\lambda}$	 			 &  0  		& 2			\\
            $\apvec{i}{\chi}^\nu$			 &	0		& $\checkmark$	\\
            \rowcolor{black!30}
            $g^{\nu\lambda}$				 &  0		& $\checkmark$\\
        \end{tabular}
        \rule{0.4\textwidth}{0.2mm}
        \begin{tabular}{p{3cm}p{1.5cm}p{1.5cm}}
            $\Gamma^\alpha_{\nu\lambda}$&1&$\checkmark$\\
            \rowcolor{black!30}
            $\{^\alpha_{\nu\lambda}\}$ &1& $\checkmark$\\
            $\gamma^\alpha{}_{\nu\lambda}$ &1&$\checkmark$\\
            \rowcolor{black!30}
            $\Lambda^\alpha{}_{\nu\lambda}$&1&$\checkmark$\\
        \end{tabular}
        \rule{0.4\textwidth}{0.2mm}
        \begin{tabular}{p{3cm}p{1.5cm}p{1.5cm}}
            $\chi_{\nu\lambda}$&1&$\checkmark$\\
            \rowcolor{black!30}
            $\epsilon_{\nu\lambda}$&1&$\checkmark$\\
            $\eta_{\nu\lambda}$&2&$\checkmark$\\
            \rowcolor{black!30}
            $\kappa_{\nu\lambda}$&2&$\checkmark$\\
        \end{tabular}	
        \rule{0.4\textwidth}{0.2mm}
        \begin{tabular}{p{3cm}p{1.5cm}p{1.5cm}}
            $\phi_{\nu\lambda}$&2&$\checkmark$\\
            \rowcolor{black!30}
            $\psi_{\nu\lambda}$&1&$\checkmark$\\
            $\varpi_{\nu\lambda}$&2&$\checkmark$\\
            \rowcolor{black!30}
            $\omega_{\nu\lambda}$&2&$\checkmark$\\
        \end{tabular}
        \rule{0.4\textwidth}{0.2mm}
        \begin{tabular}{p{3cm}p{1.5cm}p{1.5cm}}
            $R_{\nu\lambda}$&1&$\checkmark$\\
            \rowcolor{black!30}
            $F_{\nu\lambda}$&1&$\checkmark$\\		
        \end{tabular}	
        %%\begin{tabular}{p{6cm}}
            %%\checkmark: means 2 and higher powers
        %%\end{tabular}
    \end{table}

    The symmetric part of Eq.\eqref{eqn6-1::5w} will be expanded to order linear in the perturbation parameter s. Let us rewrite Eq.\eqref{eqn6-1::5w} in the following equivalent form.
    
    \begin{equation}
    \begin{split}
    R_{\nu\lambda}=-{1\over2}\big(\apvec{g}{T}_{\nu\lambda}-{1\over2}g_{\nu\lambda}\apvec{g}{T}\big).
    \end{split}
    \end{equation}
    
    %\subsection{Linearizing the Ricci Tensor}
    The following gives the s-linearized Ricci tensor.
    \begin{equation}
    \begin{split}
    R_{\nu\lambda}&=R^\alpha{}_{\nu\lambda\alpha}\\
    &=\{^\alpha_{\nu\alpha}\}_{,\lambda}-\{^\alpha_{\nu\lambda}\}_{,\alpha}+\{^\epsilon_{\nu\alpha}\}\{^\alpha_{\epsilon\lambda}\}-\{^\epsilon_{\nu\lambda}\}\{^\alpha_{\epsilon\alpha}\}\label{eqnLina},
    \end{split}
    \end{equation}
    where using the result of Eq.\eqref{eqnLin:f}, we obtain the following.
    \begin{equation}
    \begin{split}
    \{^\alpha_{\nu\lambda}\}_{,\lambda}&={s\over2}\big(W_{\nu\alpha,\alpha}+W_{\alpha\alpha,\nu}-W_{\nu\alpha,\alpha}\big)_{,\lambda}\\
    &={s\over2}W_{\alpha\alpha,\nu\lambda}\label{eqnLinb}
    \end{split}
    \end{equation}
    \begin{equation}
    \begin{split}
    \{^\alpha_{\nu\lambda}\}_{,\alpha}&={s\over2}\big(W_{\nu\alpha,\lambda}+W_{\lambda\alpha,\nu}-W_{\nu\lambda,\alpha}\big)_{,\alpha}\\
    &={s\over2}\big(W_{\nu\alpha,\lambda\alpha}+W_{\lambda\alpha,\nu\alpha}-W_{\nu\lambda,\alpha\alpha}\big)\label{eqnLinc}.
    \end{split}
    \end{equation}
    Terms involving products of the connections will have quadratic power of the parameter s and will be set to zero in this linearization scheme. So, we obtain the linearized Ricci tensor to be:
    \begin{equation}
    \begin{split}
    R_{\nu\lambda}={s\over2}\big(W_{\alpha\alpha,\nu\lambda}+W_{\nu\lambda,\alpha\alpha}-W_{\nu\alpha,\lambda\alpha}-W_{\lambda\alpha,\nu\alpha}\big)\label{eqnLind}.
    \end{split}
    \end{equation}    
    %\subsection{Linearizing the Geometric Energy-Momentum tensor}
    Next, we s-linearize the geometric energy-momentum tensor given at Eq.\eqref{eqn6-1::5v}.
    From the results of Table \ref{tab6-1::3}, we see that the only term linear in s will be $\psi_{\nu\lambda}$. So neglecting nonlinear powers of s, we have the linearized geometric energy-momentum tensor as:    
    \begin{equation}
    \begin{split}
    \apvec{g_*}{T}_{\nu\lambda}=-q^3s\psi_{\nu\lambda}.
    \end{split}
    \end{equation}
    Hence, the linearized symmetric part of Eq.\eqref{eqn6-1::5w} is:
    \begin{equation}
    \begin{split}
    {s\over2}\big(W_{\alpha\alpha,\nu\lambda}+W_{\nu\lambda,\alpha\alpha}-W_{\nu\alpha,\lambda\alpha}-W_{\lambda\alpha,\nu\alpha}\big)=-q^3s\psi_{\nu\lambda}\label{eqnLing}.
    \end{split}
    \end{equation}
    For linearized GR, the only non-vanishing term arising due to speeds in the classical regime $(v<<c)$, is the term for which the Riemannian connection is $\{^\alpha_{00}\}$; this means our linearization scheme will involve terms for which $\nu=\lambda=0$. And also, because of the low speeds involved in this region, the field must be weak and therefore assumed static so that $g_{\nu\lambda,0}=0$; this implies also that $\apvec{i}{\chi}_{\nu}{}_{,0}=0$ and consequently $W_{\alpha\alpha,0}=W_{0\alpha,0}=0$. We may therefore write Eq.\eqref{eqnLing} as:
    \begin{equation}
    \begin{split}
    {1\over2}\big(W_{\alpha\alpha,00}+W_{00,\alpha\alpha}-W_{0\alpha,0\alpha}-W_{0\alpha,0\alpha}\big)=-q^3\psi_{00}.
    \end{split}
    \end{equation}
    The terms involving derivative along the zeroth (or time) axis vanishes because we are in the static field domain. Summing over $\alpha$ and noting that the zeroth component will also yield static result, we have the following 3-dimensional second order differential equation with a source term.
    \begin{equation}
    \begin{split}
    W_{00,jj}={\partial\over\partial x^j}{\partial\over\partial x^j}\Big(W_{00}\Big)=-2q^3\psi_{00};\quad j\in\{1,2,3\}\label{eqnLini}.
    \end{split}
    \end{equation}
    In the weak field limit, we posit that the classical gravitational potential is proportional to the (linearized) gravitational potential $g_{\nu\lambda}$. That is, we may use the result of Eq.\eqref{eqnLin:e} and write:
    \begin{equation}
    \begin{split}
    \phi(x)\propto \delta_{\nu\lambda}+sW_{\nu\lambda}+O(s^2)\label{eqnLinj}.
    \end{split}
    \end{equation}
    Eq.\eqref{eqnLinj} implies:
    \begin{equation}
    \begin{split}
    W_{00}={\phi\over sk}-{1\over s}\label{eqnLink}
    \end{split}
    \end{equation}
    where k is constant of proportionality. Now, we may write Eq.\eqref{eqnLini} as:
    \begin{equation}
    \begin{split}
    \nabla^2\phi=-2q^3sk\psi_{00}=\tau\psi_{00}\label{eqnLinl}
    \end{split}
    \end{equation}
    
    Let us note that the energy-momentum distribution for Einstein's GR has been modeled on a special case of the fluid solution. This model is called the perfect fluid and it is given by:
    \begin{equation}
    \begin{split}
    T_{\nu\lambda}=\big(\rho+p\big)U_{\nu}U_{\lambda}+pg^\prime_{\nu\lambda};
    \end{split}
    \end{equation}    
    where $g^\prime_{\nu\lambda}=diag(-1,1,1,1)$, $\rho$ is the energy density of the fluid and p is the fluid pressure. In GR, the universe is modeled on this fluid model where $\rho$ is taken to be the density of energy distribution and p is the pressure within the material or source of gravity\cite{Ioannis1997}. In this linearization scheme, we have seen that we must pay attention to the case for which $\nu=\lambda=0$; recalling that $U_0=1, g^\prime_{00}=-1$, then it is easy to see that for this case, we shall have:
    \begin{equation}
    \begin{split}
    T_{00}=\rho+p-p=\rho.
    \end{split}
    \end{equation}
    And linearized Einstein's GR may be written in terms of the energy density as:    
    \begin{equation}
    \begin{split}
    \nabla^2\phi=\tau\rho\label{eqnLino}
    \end{split}
    \end{equation}
    Eq.\eqref{eqnLino} is the Newtonian limit of Einstein's GR and clearly it corresponds with the geometric result of Eq.\eqref{eqnLinl}. Hence, the symmetric part of the combined field equations has a well behaved linear approximation. To ascertain whether the present theory effectively unifies gravity and electromagnetism, the linearization scheme is also applied to the skew-symmetric part of the field equations and the result compared with the Maxwell equations.
    %\subsection{Linearizing the skew-symmetric part of the field equations}
    %%%Let us recall equation \eqref{eqn6.6a}, the skew-symmetric part of the field equations.
    %%%\begin{equation}
    %%%\begin{split}
    %%%F_{\nu\lambda}=q\big(C_{\nu,\lambda}-C_{\lambda,\nu}\big)\label{eqnskew1a}
    %%%\end{split}
    %%%\end{equation}
    %%%%Where we also recall that we set as the electromagnetic field tensor, the following.
    %%\begin{equation}
    %%\begin{split}
    %%(1-q)\kappa_{\nu\lambda}+q\eta_{\nu\lambda}=F_{\nu\lambda} \label{eqnskew1b}
    %%\end{split}
    %%\end{equation}
    
    Our next task is to linearize Eq.\eqref{eqn6.6a}. Using the results of Table \ref{tab6-1::3}, we see that $\kappa_{\nu\lambda}$ and $\eta_{\nu\lambda}$ have no terms constant nor linear in s. This implies that setting to zero terms of quadratic or higher order of s, we have $C_{\nu,\lambda}=C_{\lambda,\nu}$ for all values of $\nu$ and $\lambda$. This means that $F_{\nu\lambda}$ of Eq.\eqref{eqn6.6a} is identically zero and thus, cannot represent the electromagnetic field tensor of Maxwell theory.
    
    \section{Remarks}
    We have shown that the Dolan-McCrea variational  method, which has been underrepresented over the years is a potent tool for Physics by applying it extensively and analytically in deriving combined field equations of gravity and \textquotedblleft electromagnetism\textquotedblright. The gravity component of the nonsymmetric equations behaved as expected up to linearity. However, the electromagnetism component vanishes identically at linearity. We state here that this unexpected behavior has been attributed to the invariant scalar used in the variational method and a scalar has been constructed from the W-tensor to address this issue\cite{Wanas2015}.
       
    Also, we have dotted the i's and crossed the t's while lending some physical significance to geometric results.

    \section*{Acknowledgement}
    %We are thankful to Consejo Nacional de Ciencia y Tecnología (CONACyT) of México  and to Centro de Innovación y Desarrollo Tecnológico %en Cómputo (CIDETEC), IPN. Also, many Thanks to the COMSATS Institute of Information Technology, Islamabad.
    We are thankful to the following: the National Mathematical Center (NMC), Abuja, the COMSATS Institute of Information Technology (CIIT), Islamabad and the Higher Education Commission (HEC) of Pakistan for jointly funding our research.
    %%%%%%%%%%%%%%%%%%%%%%%%%%%%%%%%%%%%%%%%%%%%%%%%%%%%%%%%%%%%%%%%%%%%%%%%%%%%%
    %\bibliographystyle{aipauth4-1}
    \bibliographystyle{apsrev4-1}
    \bibliography{bibfile}
    
\end{document}